# A curious case of the Indian Summer Monsoon 2020: The influence of Barotropic Rossby Waves and the monsoon depressions.

Nimmakanti Mahendra[1], Nagaraju Chilukoti[1,*], Jasti S Chowdary[2], Ashok Karumuri[3]

Manmeet Singh[2]

[1] Department of Earth and Atmospheric Sciences, National Institute of Technology, Rourkela – 769008, India.

[2] Indian Institute of Tropical Meteorology, Ministry of Earth Sciences, Pune - 411 008, India.

[3] Centre of Earth, Ocean and Atmospheric Sciences, University of Hyderabad, Hyderabad, India

*Email: chilukotin@nitrkl.ac.in/chilukotinagaraju@gmail.com

**Abstract**

The seasonal summer monsoon rainfall over India has substantially depended on the synoptic-scale systems such as monsoon lows and depressions. India has received above-average rainfall during the 2020 summer monsoon season. Total 12 Low Pressure Areas (LPAs) formed in the north Indian Ocean during summer 2020 (in JJAS season). The significance of this monsoon season is that August 2020 received the highest all-India rainfall in the past 44 years since 1976. This is accompanied by around 50% of the total seasonal LPAs formed in August 2020, none of which intensified into a monsoon depression (MDs). This study attempts to understand the characteristic features of monsoon rainfall during August 2020 and explore the plausible mechanisms behind the LPAs not intensifying/concentrating as MDs. It is noted that the anomalous warming over the Northern Parts of the Arabian Sea (NPAS) resulted in increased convection over this region in August 2020, as a result, strong convergence of low-level wind is observed over NPAS region. In addition to this convergence, strong northwesterly winds emanating from central Asia merged with the enhanced cross-equatorial monsoon flow. However, this strong flow over the Arabian Sea sheared/dissociated into two branches: one extending up to northwest (NW) India along the monsoon trough, another one diverging into an anticyclone over the south BOB (SBOB), which reduced the horizontal shear there (Barotropic Instability). This anticyclone strength over the SBOB and its westward shift is determined by the western north pacific (WNP) anticyclone. Our analysis suggests that due to the poor barotropic instability over the head BOB, LPAs could not develop into MDs.

Additionally, upper level (200 hPa) barotropic Rossby wave in August 2020 remains stationary over South Central Asia and retrogressed with a northeast to southwest orientation. It determined the path of movement of the low-level disturbance beneath and affected the all-India rainfall by virtue of enhanced rainfall over NW & Western Ghats (WG) regions. The interplay of the barotropic Rossby wave alongside an anticyclone over the WNP accompanied by local conditions caused the above normal rainfall over India in August 2020, even though there are adverse dynamical conditions. We have also verified these mechanisms in Community Earth System Model Large Ensemble (CESM-LE) model simulations, Analysis shows that model has a limited skill to simulate the changes in the monsoon rainfall and associated circulation and failed to capture the mid-latitude circulation impact on ISM rainfall.

## 1. Introduction

The seasonal cycle of the Indian summer monsoon (ISM) rainfall exhibits month-to-month variations in recent years, mainly in the peak monsoon months (July and August). These months contribute more than 60 to 70 % of the annual rainfall and are directly associated with the country's agriculture and socio-economic activities. Therefore, it is necessary to explore adequate mechanisms for month-to-month variation in ISM rainfall in the changing climate to understand the monsoon habitude for the improvement of predictions. ISM rainfall is significantly modulated by synoptic-scale systems such as Monsoon depressions (MDs) and monsoon lows/Low Pressure areas (LPAs). LPAs and MDs, which normally originate in the Bay of Bengal north of 18°N latitude and move in a west-northwesterly direction across the central and northern parts of India, provide plentifully rainfall over the region where they form and move (Koteswaram and Rao, 1963; Krishnamurthy and Ajayamohan,2010), lasting an average of 4-5 days (Mooley 1973; Krishnamurti et al. 1975). According to the Indian Meteorological Department (IMD), MDs are said to occur when two closed isobars (at 2hPa intervals) can be drawn on the surface synoptic weather charts and winds in the cyclonic circulation are between 17 to 33 knots. Their shape is roughly elliptical with 1000's kms of horizontal and 9-6 km vertical extension. These are cold core systems with warm core aloft. The MDs tilt southwards with height and if they move westward, the heavy rainfall is mainly concentrated in the southwest(SW) quadrant. Due to high vertical wind shear present during the SW monsoon season generally MDs do not intensify into cyclonic storms. Weaker systems with only one closed isobar on the surface pressure chart (at 2hPa intervals) must be within 3° of horizontal extension and wind speed not exceeding 17 knots are called LPAs (Hunt et al. 2019). The number of MDs has been decreasing significantly in recent decades due to low mid-tropospheric relative humidity (e.g., Vishnu et al., 2020); however, Praveen et al. (2015) attributed observed 60% of monsoon rainfall in India was associated with LPAs. Further, Hunt et al. (2019) quantified that in recent years most synoptic rainfall was received in India by short-lived LPAs. They determined that, the rainfall contributed by LPAs (MDs) are nearly 57%(17%) over the monsoon core region and 44%(12%) overall in India. Likewise, India received above-average monsoon rainfall in 2020, especially during August, with around 50% of the total seasonal LPAs formed during this month (12 LPAs formed during 2020 summer monsoon season). An essential feature of the 2020 summer monsoon is that out of 12 LPAs, none of them intensified into a depression. Hence, it is necessary to understand the physical mechanisms responsible for the failure of developing into MDs in recent decades. Previous studies have well documented the decline of MDs in recent decades, but those are not fully-understood reasons behind the failure of the MDs (e.g., Vishnu et al., 2016).

In this study, we investigate the reasons behind the failure of LPAs to intensify into MDs during August 2020 and factors that are responsible for anomalously high rainfall over Indian subcontinent. A significant large-scale feature during the ISM 2020 is the development of La Niña in the Pacific Ocean (e.g., Chu et al., 2021; Pan et., 2021 and Qiao et al., 2021), co-occurring with basin-wide warming in the Indian Ocean. Numerous studies on the El Niño Southern Oscillation (ENSO) -monsoon relationship indicate that the ISM rainfall is significantly modulated by the teleconnections between the air-sea interaction processes of global oceans (e.g., Sikka 1980; Pant and Parthasarathy 1981). ENSO significantly influences ISM rainfall in the inter-annual time scale (Webster and Yang 1992; Kripalani and Kulkarni 1997; Chowdary et al. 2015; Ashok et al. 2014; Nagaraju et al. 2018; Singh et al. 2020; Mahendra et al. 2021). The La Niña (cold phase of ENSO) generally induces positive or excessive rainfall anomalies over India (e.g., Sikka 1980; Pant and Parthasarathy 1981; Bhalme et al. 1983; Webster and Yang 1992; Yadav 2009a; Singh et al. 2020). Apart from the tropical Pacific, the Tropical Indian Ocean (TIO) also plays a significant role in ISM rainfall (e.g., Yang et al. 2009; Chowdary et al. 2017). The TIO has been warming rapidly in recent years with the changing climate and acting as an independent force on monsoons, like the ENSO and aerosols (e.g., Webster et al., 1999; Saji et al., 1999; Ashok et al., 2001, Fadnavis et al. 2020). Perhaps, it has reduced the influence of ENSO on ISM in recent decades, but the ENSO, Indian Ocean Dipole (IOD) relationship also decreased (Ham et al., 2017). The IOD is a natural ocean-atmosphere coupled phenomenon of the TIO which exerts significant influence on the regional and global climate (Behera et al., 1999; Saji et al., 1999; Webster et al., 1999; Black et al., 2003; Yamagata et al., 2004; Kripalani et al., 2005). The IOD corroborate by Western North Pacific (WNP) low-level circulation modulating the teleconnections of ISM rainfall regionally (e.g., Mahendra et al. 2021); these changes in the Indo-Pacific Ocean are well documented in recent decades (e.g., Graham, 1994; Trenberth and Hurrell, 1994). Positive IOD can modulate rainfall over central India by supporting the enhanced genesis of MDs over the BOB (e.g., Ajayamohan and Rao 2008). Krishnan et al. (2006) reported that Positive IOD-induced changes in circulation and dynamics reinforce the activity of the monsoon MDs by strengthening cross-equatorial moisture transport from the south-eastern tropical Indian Ocean (SETIO) into the BOB. It enriches the barotropic instability of monsoon flow (Behera et al., 1999; Ashok et al., 2004) and favours the MDs.

The Tibetan High (TH) is a semi-permanent system during the ISM. The transit of the Jet-stream over the Indian region amplifies the role of the TH and thereby induces an active southwest monsoon (e.g., Krishnamiurti and Bhalme 1976). The jetstream plays as a waveguide and causes the appearance of a frequent Rossby wave. This kind of intertwining of Rossby wave propagation along with the jetstream is known as the Silk Road Pattern (SRP). The wave shears off near the exit region of the jet stream, and it is called "Rossby wave breaking". This breaking induces huge potential vorticity in the region, influencing the ISM (Takemura et at., 2020). Joseph and Srinivasan (1999), found that it shows different spatial phases between dry and wet Indian monsoon years. Yadav (2017) study yielded that the Eurasian Rossby wave train, known as the SRP, has positive feedback on ISM circulation giving rise to anomalous north-westerly winds over central Asia merging with mean monsoon flow in the Arabian Sea. Pick up the moisture from the warm Arabian Sea and the Persian Gulf for the deep convection over the Western Ghats, north-west, and central India. Hence, it reinforces active ISM conditions. Ding et al. (2011) suggested that the ENSO could also excite the SRP and regulate its phases. The studies by Hoerling and Kumar (2003) and Lau et al. (2005) documented that the Indo–western Pacific SST warming could generate a belt of positive, zonally symmetric, upper-level geopotential height that could induce anomalous warmth and dryness. This suggests that the barotropic Rossby wave pattern is closely associated with Asian summer monsoon rainfall (ISM) variability and is also influenced by concurrent ENSO signals. Therefore, one needs to understand how changes in the Indo-Pacific climate influence the mid-tropospheric Rossby wave pattern and how it affects the characteristics of monsoon synoptic features by modulating the mid-latitude Rossby wave during August 2020. The present paper endeavours to analyze these aspects of August 2020. The manuscript is organized as follows: data and methodology are discussed in section 2, results are discussed in section 3, and summary & conclusions are discussed in section 4.

**2. Data and Methodology**

In this study, the Sea Surface Temperature (SST) data from Hadley Center Sea Ice and SST version 1.1 (HadISST; Rayner et al., 2003) is used for the period of 1870 to 2021. HadISST data was reconstructed from observational data and obtained from the buoy, ships, satellite, and remote sensing, with the resolution of the dataset 1°x1°. We also used India Meteorological Department (IMD) gridded rainfall data (0.25°x0.25°; Rajeevan et al. 2006) from 1951 to 2020 in this study. GPCP precipitation data (2.5°x2.5°) is taken from NCEP/NCAR. All the other meteorological parameters such as winds (zonal and meridional), specific humidity, relative humidity and Mean Sea level pressure data are obtained from the ERA5 reanalysis monthly (0.25°x0.25°) high-resolution data (Hersbach et al., 2019). We have also used Community Earth System Model Large Ensemble (CESM-LE) model simulation data sets which are freely downloadable available from the website https://www.cesm.ucar.edu/projects/community-projects/LENS/data-sets.html (Kay et al. 2015) to evaluate the model fidelity in simulating observed changes in these physical mechanisms. The frequency of Monsoon Depressions is taken from the Cyclone e-Atlas-IMD data sets which can be downloaded from website http://www.imdchennai.gov.in/cyclone_eatlas.html during the period 1979 through 2020 and the data of LPAs are gleaned from the weekly weather reports issued by the IMD. Monthly anomalies for all the datasets are constructed by removing their respective climatology. The seasonal rainfall departure of 2020 was calculated (as per the IMD criteria) using climatology for the period of 1951 to 2020. The monsoon rainfall during August can be statistically evaluated and interpreted by means of an index called Standardized Rainfall Index-August (SRI-A; for monsoon rainfall in August). The index takes into account the area-weighted monsoon rainfall over central India and Northwest & Western Ghats (NW & WG) for a considerable period of time. The inter-annual monsoon rainfall is graphed by means of standardized anomalies. The SRI-A standardizes the anomalies as the number of standard deviations (SD) from the long period monsoon August rainfall.

$$SRI\text{-}A = (x - Mean)/SD$$

Rossby wave activity was calculated based on Takaya and Nakamura (2001) formulas as given below:

$$W_x = \frac{p \cos \varphi}{2|U|} \left( \frac{U}{a^2 \cos^2 \varphi} \left[ \left(\frac{\partial \psi'}{\partial \lambda}\right)^2 - \psi' \frac{\partial^2 \psi'}{\partial \lambda^2} \right] + \frac{V}{a^2 \cos \varphi} \left[ \frac{\partial \psi'}{\partial \lambda} \frac{\partial \psi'}{\partial \varphi} - \psi' \frac{\partial^2 \psi'}{\partial \lambda \partial \varphi} \right] \right)$$

$$W_y = \frac{p \cos \varphi}{2|U|} \left( \frac{U}{a^2 \cos \varphi} \left[ \frac{\partial \psi'}{\partial \lambda} \frac{\partial \psi'}{\partial \varphi} - \psi' \frac{\partial^2 \psi'}{\partial \lambda \partial \varphi} \right] + \frac{V}{a^2} \left[ \left(\frac{\partial \psi'}{\partial \varphi}\right)^2 - \psi' \frac{\partial^2 \psi'}{\partial \varphi^2} \right] \right)$$

Here, W is wave flux in the horizontal (Wx) and meridional (Wy) directions, p is pressure normalized by 100hPa, (φ, λ) represents latitude and longitude. U, V are zonal and meridional winds respectively, and |U| is magnitude. Geopotential anomalies will be used to compute perturbation stream-function ψ' in Quasi-Geostrophic (QG) assumption (ψ'=φ−φ/f), where φ is geopotential height. "f" is the Coriolis parameter: f=2Ωsinϕ.

## 3. Results and discussions

The spatial rainfall distribution during the 2020 ISM season (Fig. 1b) shows large spatial variability over India as compared with climatology (Fig 1a). The country received an above-normal monsoon during the 2020 summer monsoon (Fig 1c). The country received 109 % rainfall of the long period average, with August witnessing above normal rainfall, while July recorded deficient rainfall in Central India, which is evident in Figure 1d.

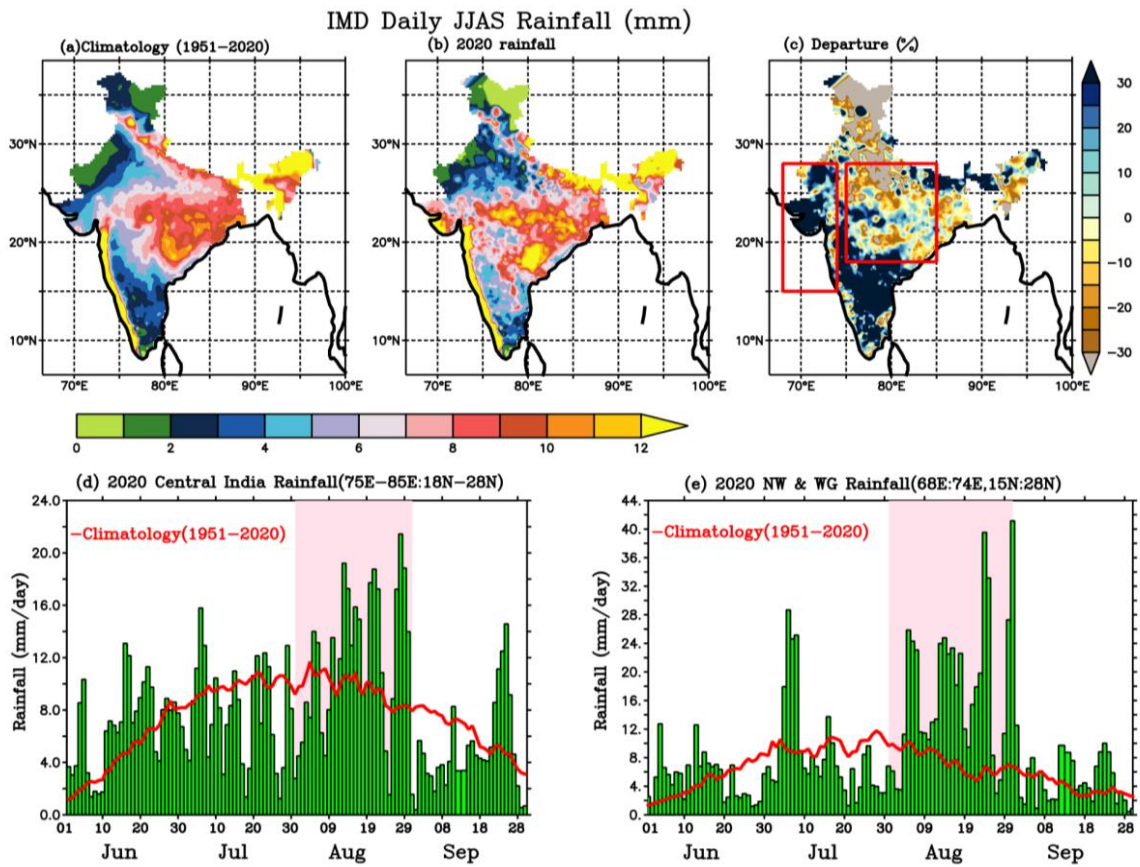

**Figure 1:** Spatial-temporal distribution of 2020 summer monsoon depicting daily rainfall, (a) climatology rainfall (mm/day) for the period of 1951-2020, (b) 2020 JJAS rainfall (mm/day), (c) rainfall departure (%) during 2020, calculated (as per the IMD criteria) using climatology for the period of 1951-2020. (d) Bars represent Central India (75°E:85°E-18°N:28°N) daily rainfall during 2020, overlaid (red line) climatology over Central India. (e) same as (d) for Northwest and Western Ghats (68°E:74°E-15°N:28°N), August month highlighted in pink shade.

Excess (deficit) monsoon rainfall increases (decreases) crop productivity which in turn affects farmer income and the country's economy (e.g., Gadgil and Rupa Kumar 2006). August 2020 received the highest all-India rainfall in the last 44 years since 1976

(https://www.imdpune.gov.in/Clim_Pred_LRF_New/Reports/). However, negative departures were seen over parts of the monsoon core region (Central India) in 2020 (fig. 1c); this can be partly due to the deficiency of MDs. MDs are declining significantly in the last three decades (e.g., Vishnu et al., 2016) which is evident in figure (2a). Though, northwest parts of India (especially over Gujarat) show positive departure and it clearly suggests that August 2020 received all-India rainfall partly on account of heavy rainfall over NW & WG (68°E:74°E-15°N:28°N) which is evident in rainfall (fig 1e) averaged over that region.

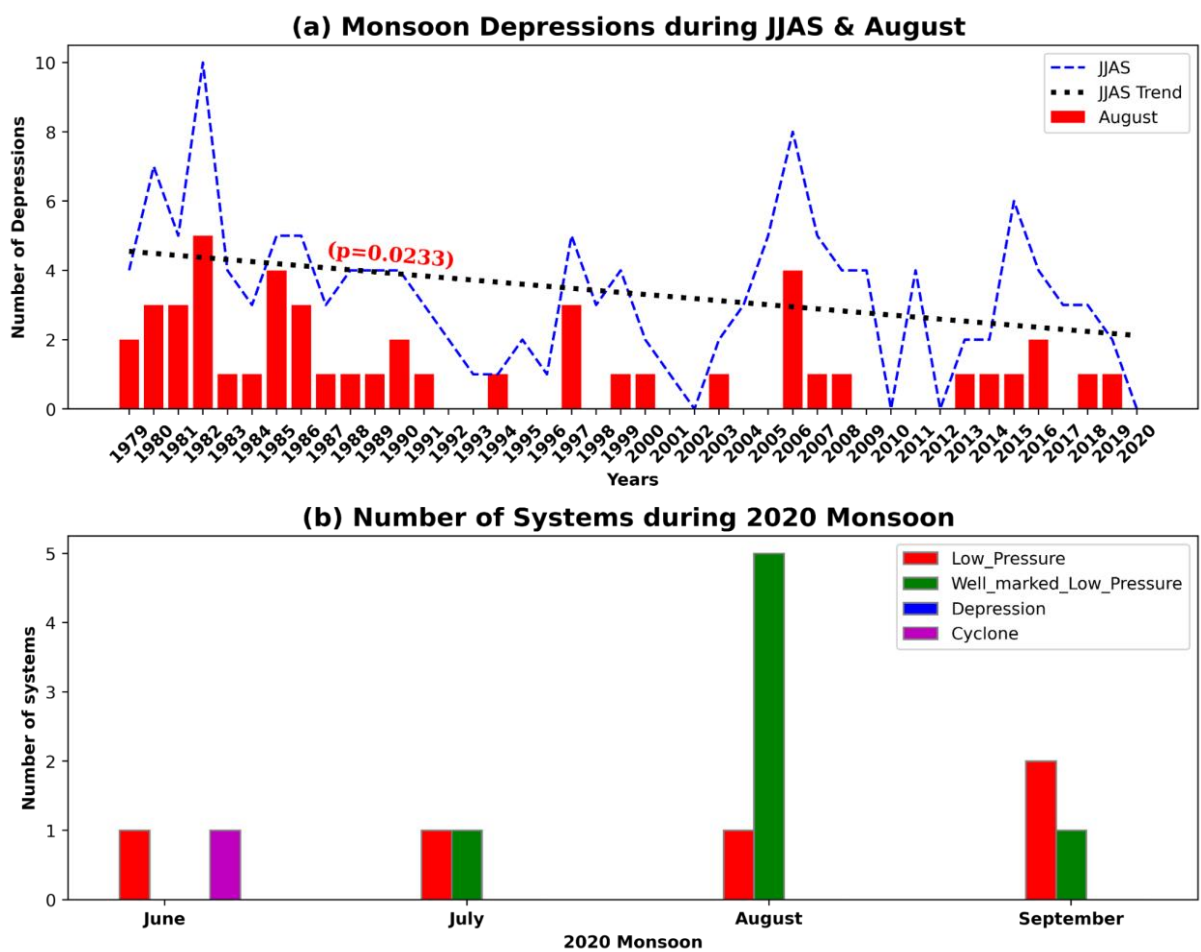

**Figure 2.** (a) The number of monsoon depressions for the period of 1979 to 2020 here dotted line represents the MDs during JJAS, black dotted line represents its trend significant at a 95% confidence level and bars show MDs during August for respective years. (b) Systems during the 2020 monsoon (Source: IMD daily weather reports): here color bars represent red low pressure, green well-marked low pressure, blue depression, and magenta cyclone.

**3.1 Synoptic-scale systems such as MDs and LPAs in August 2020**

The amount of monsoon rains is largely influenced by the frequency or number of low-pressure systems especially MDs which shed a copious amount of rainfall as they traverse from the north BOB to the north/northwest India along the monsoon trough and core monsoon zone. In 2020, the monsoon onset over Kerala is on June 1. The timely onset is embedded with cyclone "Nisarga". An important synoptic feature of the 2020 summer monsoon is the evolution of 12 low-pressure systems (fig 2b, 05 LPAs, and 07 well marked LPAs), none of which intensified into a depression. The large increase of monsoon precipitation during August 2020 was associated with 6 low-pressure systems, i.e around half of the total low-pressure systems during the entire season. Those are formed (4-10, 9-11, 13-18, 19-26, and 24-31) over the BOB and moved nearly west/northwestwards providing widespread rainfall along paths. As a result, August recorded the highest all-India rain in the last 44 years since 1976. The amount of precipitation corresponded to 127% of the long period average, which is evident over central India and NW & WG (fig 1b).

**3.2 The influence of large-scale features on monsoon synoptic condition**

We first examined the monthly evolution of SST and 850 hPa circulation to understand the abnormally high rainfall in August 2020. Month-to-month SST evolution clearly shows the presence of strong equatorial Indian Ocean warming during the pre-monsoon months (Fig. 3a, b). This basin-wide warming is continued to persist in all four summer months over TIO (Fig 3 c to f). In July 2020, an anomalous low-level anticyclone over the WNP is seen (Fig. 3d). Strong easterly wind anomalies in the southern flank of this anticyclone are extended to the north Indian Ocean. This easterly over the BOB and the Arabian Sea weakens the mean south-westerlies and warms the SST by reducing the latent heat flux released to the atmosphere, this is also true for June 2020. Anomalous north westerlies over central India induced by WNP anticyclone in July 2020 caused reduced rainfall in the monsoon trough region (e.g., Darshana et al .2022, under revision). However, by August 2020, easterlies associated with WNP anticyclone are weakened over the north Indian Ocean due to their weak strength. On the other hand, south-easterly anomalies from the equatorial Indian Ocean and south-westerlies over the Arabian Sea converged and strengthened the monsoon flow (Fig. 3e). These moisture-laden strong mean winds are favorable for good monsoon rainfall over India. Further, the anomalous northwesterly flow from central Asia confluence with the southwesterly flow, extending up to NPAS with anomalous cyclonic circulation response to underlying warm SST. Anomalously warm SST over this region corroborated by a low-pressure area caused low-level wind convergence there away from the mainland. This suggests that the enormous moisture converges over the NPAS, resulting in high rainfall over NW & WG (Fig 1e).

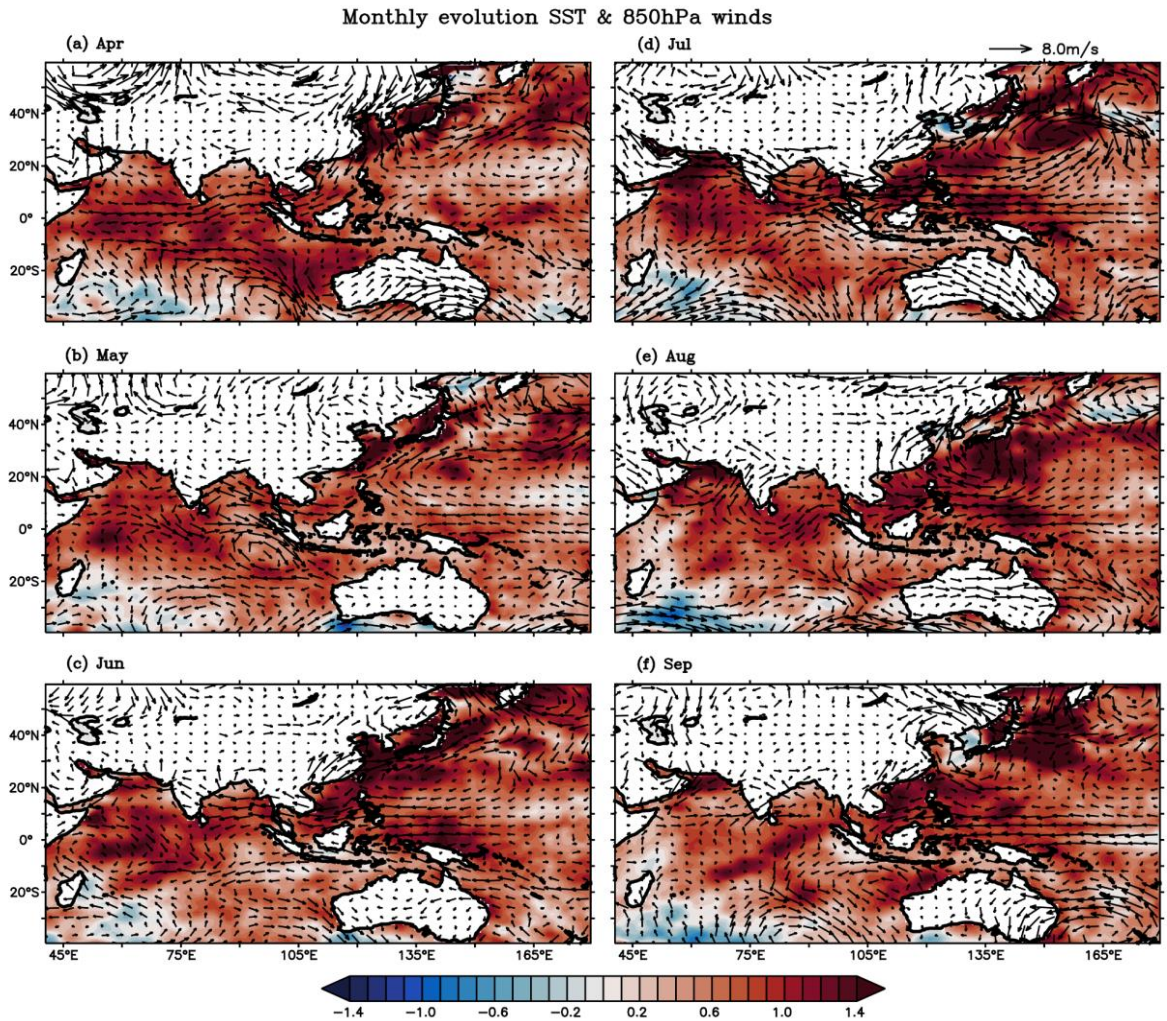

**Figure 3.** Observed monthly evolution of SST (°C, shaded) and 850 hPa winds (m/s, vectors); a to f represents each month from April to September 2020 respectively.

Further, we analyzed the circulation pattern at different levels during August 2020. Anomalous cyclonic circulation observed over NPAS is part of Gill's pattern (e.g., Gill 1980) (Fig. 4a). Strong easterlies from the east and westerlies from the west to the western equatorial Indian Ocean (WEIO) warming in the lower atmosphere are evident in figures 4a and b. The northward outflow in the lower atmosphere strengthens the cyclonic circulation over the NPAS (fig 4 d), which is seen up to 500 hPa. Strong easterlies induced by the WNP anticyclone generate the anticyclone over the southern parts of the BOB (S-BOB) (fig 4 b), which weakens the horizontal shear during August 2020, which is noticeable with a weak cyclonic circulation over head BOB. As per the above analysis, it is noted that the monsoon mean flow field is dynamically unstable. The low-level cyclonic circulation over the NPAS and anticyclone over the south BOB cut off energy from the mean flow and reduced the barotropic instability (Holton 1973). This might be one of the reasons for the failure of lows to develop into the MDs in August 2020. To authenticate this result, we compared circulation and environmental conditions of August 2020 with excess August rainfall years associated with MDs as a composite (fig 5 a and b). Excess August rainfall composite years are defined based on the central India SRI-A, which is rainfall averaged over central India (75°:85°E-18°:28°N) standardised values are greater than +1. We have selected the following years, such as 1982,1992,1994 and 2006 (fig 5 c). Figure. 5 gives a fleeting glimpse of SRI-A for the period from 1979 to 2020. It can be clearly observed from figures (5 c) that in the year 2020 August is indicating a wet monsoon month on account of NW & WG rainfall.

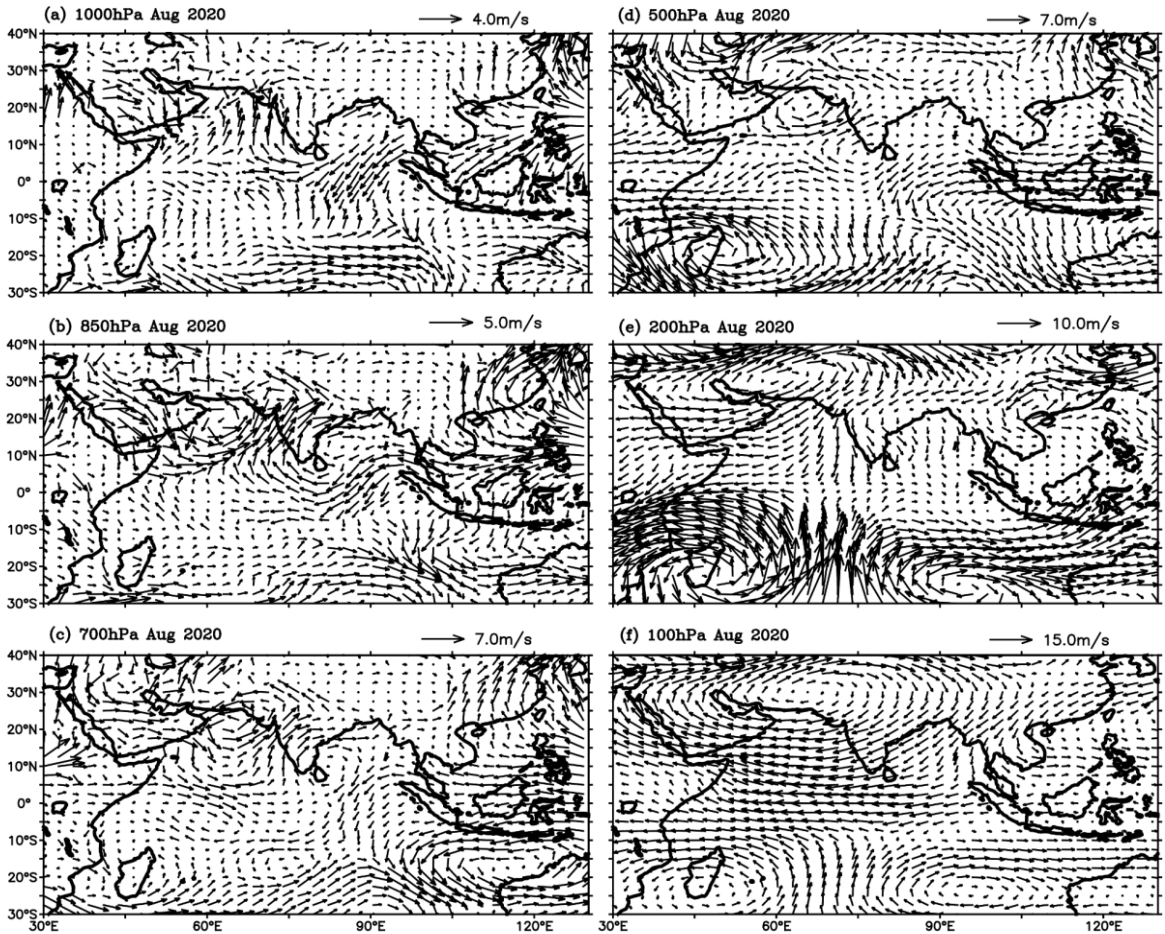

**Figure 4.** Circulation during August 2020 at different levels from (a to f) 1000, 850, 700, 500, 200, and 100 hPa respectively.

During August 2020 SST anomaly pattern clearly shows weak La Niña (fig 6 a and c) over the equatorial Pacific Ocean and it is also associated with Indian Ocean basin-wide worming (IOBW), precisely over NPAS (fig 6 c). Warming over WNP resulting from fair weather conditions prevailed due to anticyclonic circulation (e.g., Wang et al., 2021a; Qiao et al. 2021; Zhou et al. 2021; Takaya et al. 2020; Wang et al 2005; Xie et al 2001, 2002) and resulted in the shifting of east Asian monsoon convection. In the composites, a weak El Niño signal co-occurs with weak SETIO cooling (fig 6b), suggesting that the SETIO cooling is one of the important factors for the high rainfall in composites. This cooling influence on the enhancement of ISM rainfall was well documented in the previous studies (e.g., Ashok et al. 2001; Krishnan et al. 2006; Ajayamohan and Rao 2008). However, during August 2020, IOD was almost neutral, and IOBW persisted for the entire year (fig 6 c). The warming over the south equatorial TIO and NPAS is noted in fig 6 (a). This warming modified the 2020 monsoon circulation under the influence of easterlies produced by WNP anticyclone and strong north westerlies from central Asia. Note that, though most of the TIO displayed anomalous warm SST, none of the lows developed into MDs and at the same time ISM rainfall is normal during August. To understand the factors responsible for non-developing depressions, we examine essential environmental conditions for the genesis, such as relative vorticity at 950 hPa, relative humidity, etc.

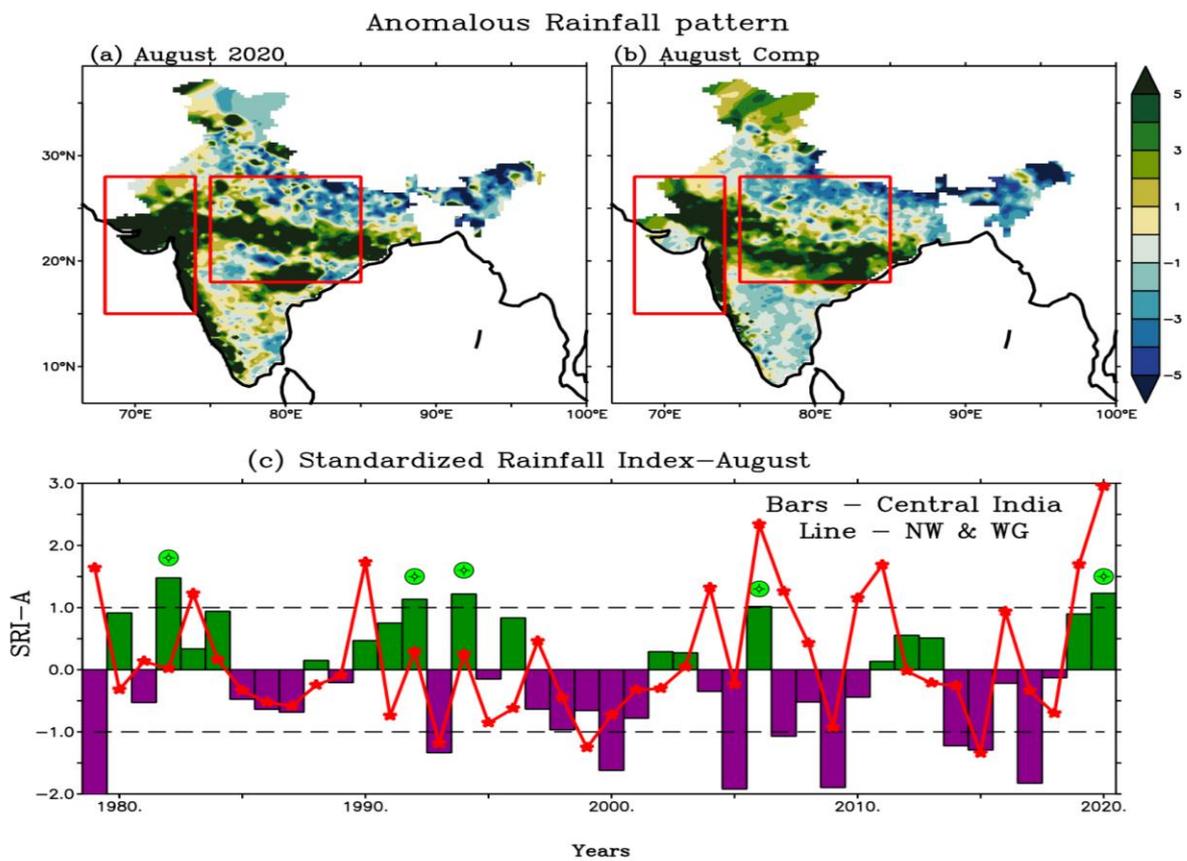

**Figure 5**. Anomalous rainfall pattern (a) during the month of August 2020, (b) August composites, and (c) standardised rainfall Index-August (SRI-A) about the period of 1979-2020, bars denote SRI-A over central Indian and time series denotes over northwest and western ghats region, here green circles denoted by SRI-A greater than 1 over central India.

The deepening of positive relative vorticity in north BOB is evident in the composite, which is weak in August 2020. Strong positive vorticity over the NPAS and an extension of negative relative vorticity from the S-BOB are seen in August 2020 (fig 7 a and b). This replicates the cyclone and anticyclone patterns over that region. One of the prominent features of the 2020 monsoon is the WNP anticyclone which displays in figures 7 (c and d), it weakens the north-south pressure gradient by producing the east-west pressure gradient during August 2020; Whereas, in the composite north-south pressure gradient is evident. Associated with this anomalous low-pressure clearly evident in 500hPa geopotential height along the monsoon trough (Fig 7 e and f) in composites, which is located over the NPAS in August 2020 associated with high-pressure extension from WNP to S-BOB. Strong warm temperature anomalies (fig 7 g and h) were observed over NW India which is absent in the composites. This suggests that during August 2020, the presence of the WNP anticyclone and anomalous SST warming over the NPAS resulted in setting-up strong the east-west pressure gradient and this might have played an important role to maintain the strong convective centre over central India. In this context, furthermore, we verified the relative vorticity over those regions such as Central India (75°:85°E - 18°:28°N), BOB (83°:93°E - 16°:23°N), Arabian Sea (60°:70°E - 18°:22°N), and S-BOB (38°:92°E - 10°:16°N).

Figure 8 shows decreased relative vorticity over the BOB in August 2020 as compared to composites and whereas, in the Arabian Sea it increased. Further increased negative vorticity over S-BOB in August 2020 is observed and which is weak in the composites (fig 8b). This negative vorticity is not favorable for the intensification of lows into MDs. Relative humidity also showed consistent results with increased in the Arabian Sea and SBOB during August 2020 as compared with composites. High relative humidity over the S-BOB during August 2020 due to large moisture transport through strong easterlies emanated by WNP anticyclone

(fig 8b). This suggests that though thermodynamic conditions are somewhat favorable, dynamics are highly influenced by large-scale circulation not favorable for depressions to grow. It is important to note that the ISM rainfall in August 2020 is above normal in spite of having no depression entering the Indian landmass.

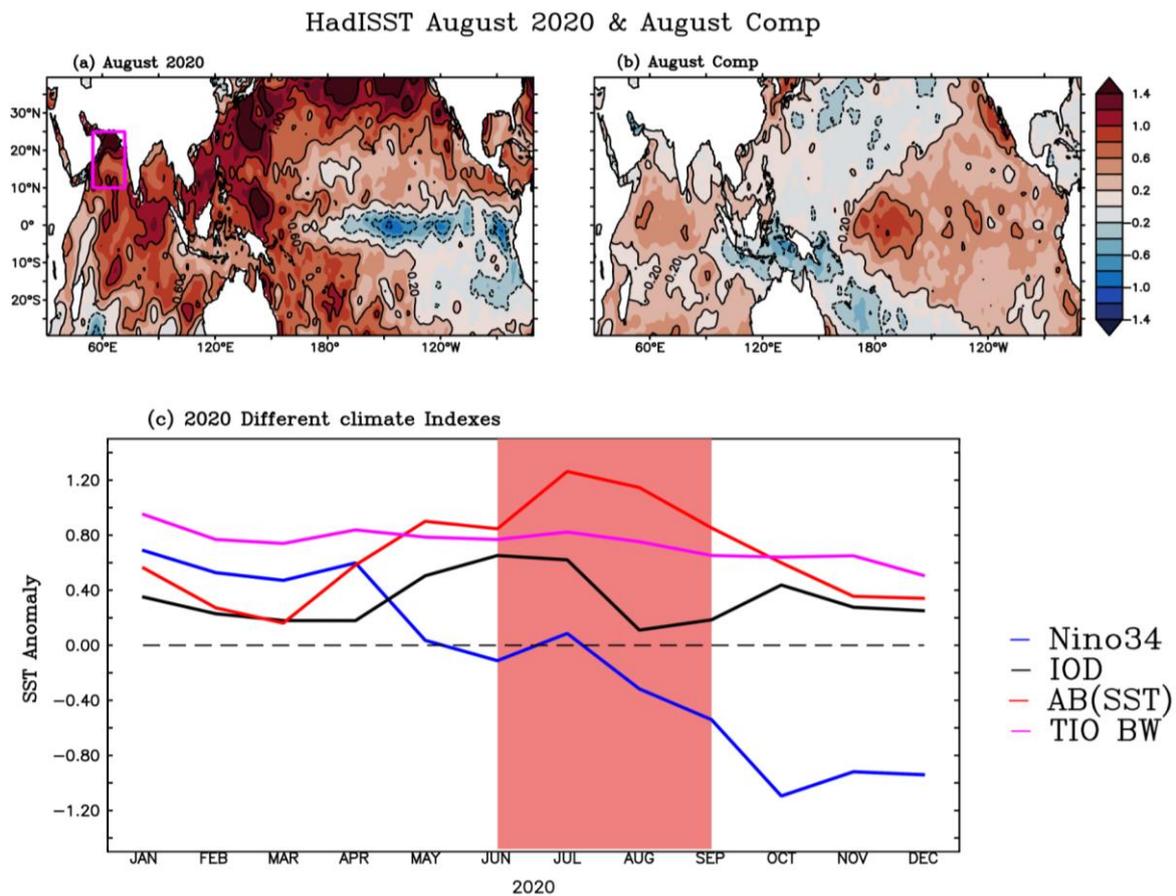

**Figure 6.** Anomalous spatial distribution of SST °C about (a) August 2020, (b) August composites, and c) representing different indices such as Indian Ocean dipole mode index, Nino34 index, averaged SST°C anomalies over northern parts of the Arabian Sea and Indian ocean basin-wide warming (TIO BW )(Pink box region indicates summer months JJAS).

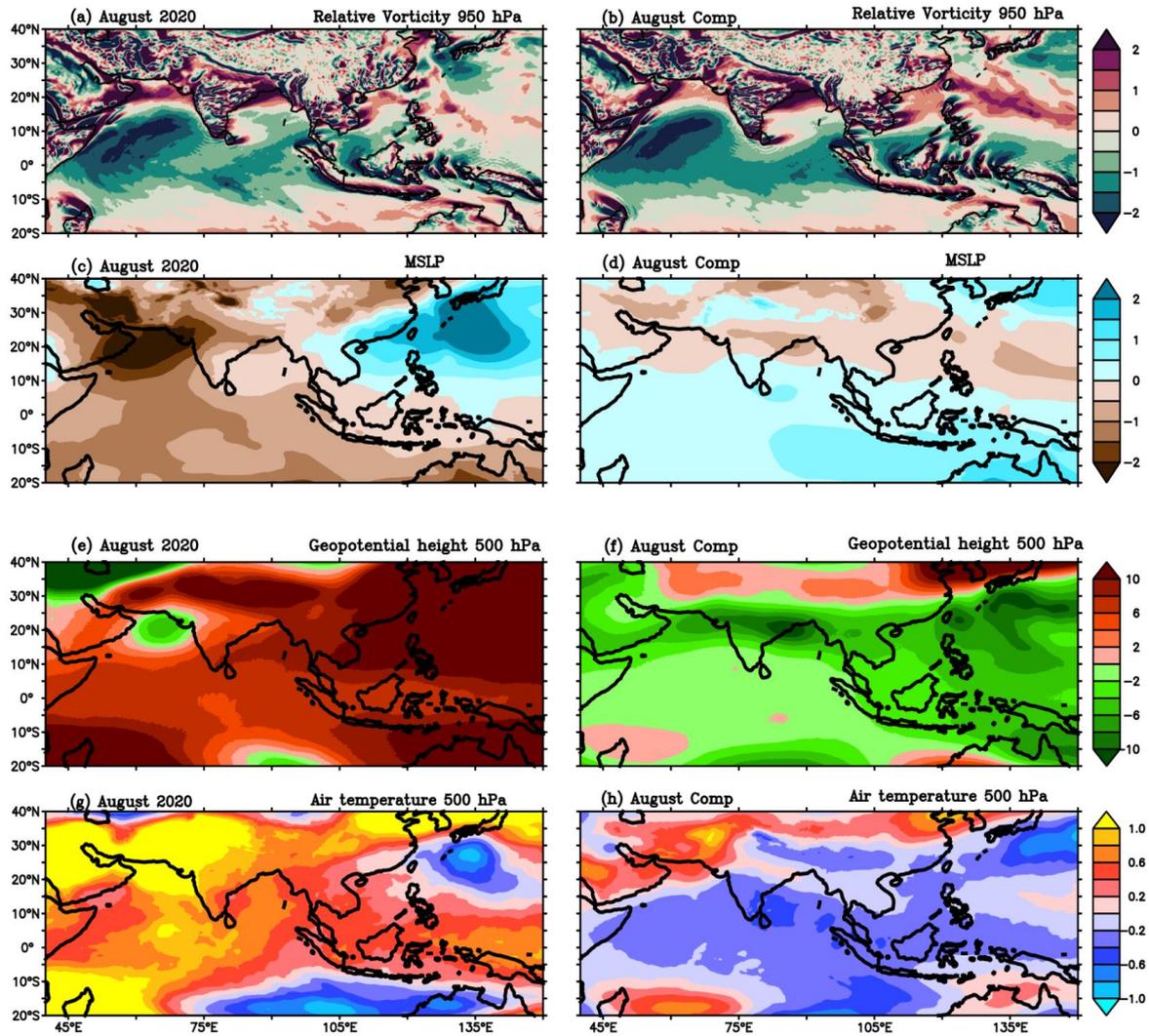

**Figure 7.** (a-b) Spatial distribution of Relative vorticity shaded (in the order of x10$^{-5}$/Sec), (c-d) Mean Sea level Pressure anomalies (hPa). (e-f) Geopotential Height (m) anomalies and air temperature (°C, g h) at 500 hPa.

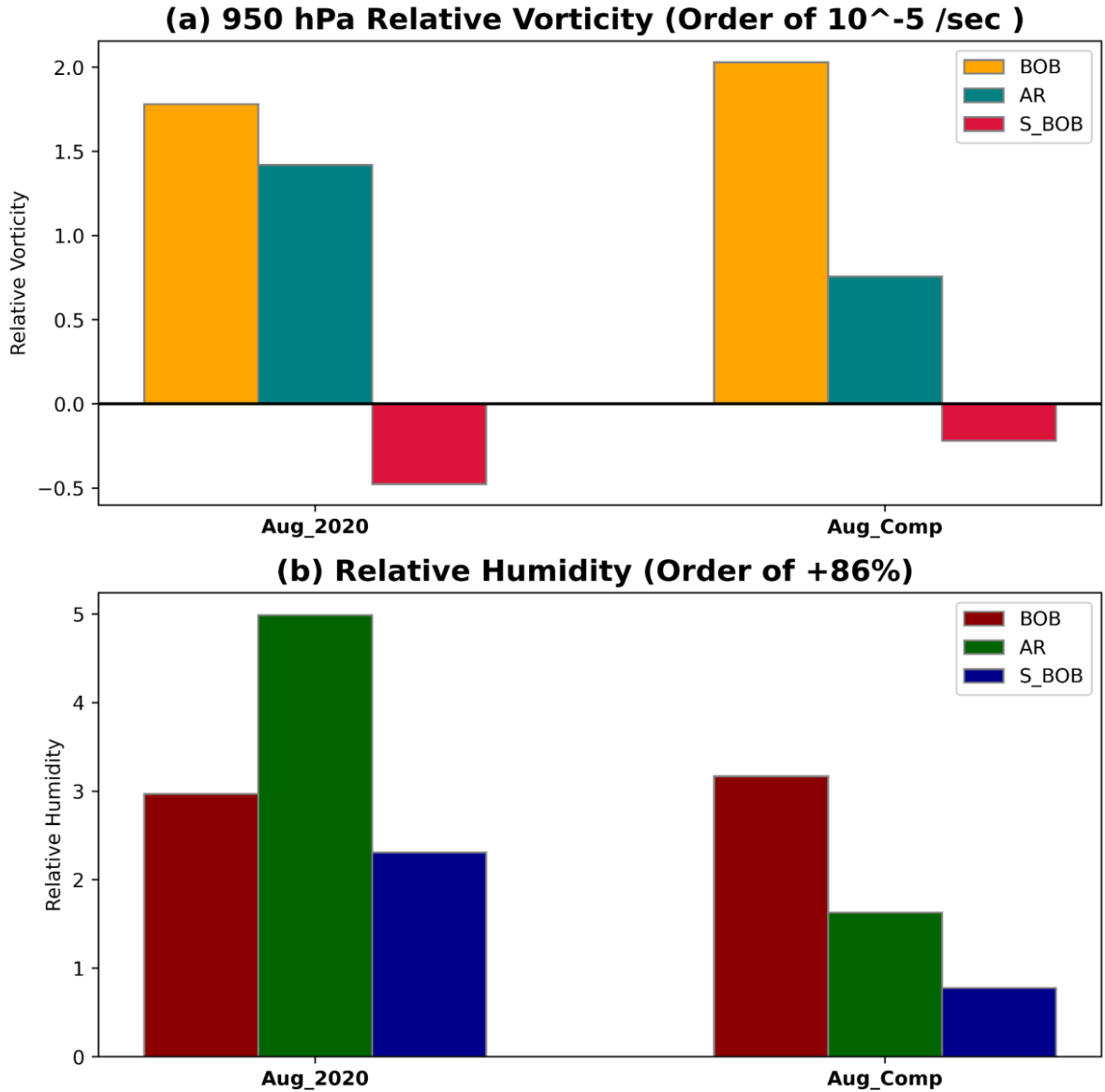

**Figure 8.** (a) Relative vorticity (in the order of x10$^{-5}$/Sec) over the regions such as the Bay of Bengal (BOB, 83°E:93°E-16°N:23°N), Arabian Sea (AB, 60°E:70°E-18°N:22°N), and South-Bay of Bengal (S-BOB, 83°E:92°E-10°N:16°N), (b) same as a but Relative humidity (Order of +86%).

Afterward, we inspect the advancement of the large-scale feature to understand the factors responsible for the above-normal rainfall during August 2020. Stream function and rotational winds at 850 hPa, and velocity potential and divergent winds at 200 hPa are displayed in Figure 9. The strong WNP anticyclone is apparently associated with the vigorous easterly flow due to its westward extension to S-BOB and resultant negative vorticity is seen over S-

BOB (fig 9a). Strong upper-level convergence over WNP and divergence over the NPAS are seen (fig 9c) which is consistent with earlier results. In the August composites, a strong cyclonic circulation is evident (fig 9 b) along the monsoon trough region associated with strong cross-equatorial flow from SETIO with weak anticyclone (fig 8b) over S-BOB as compared with August 2020.

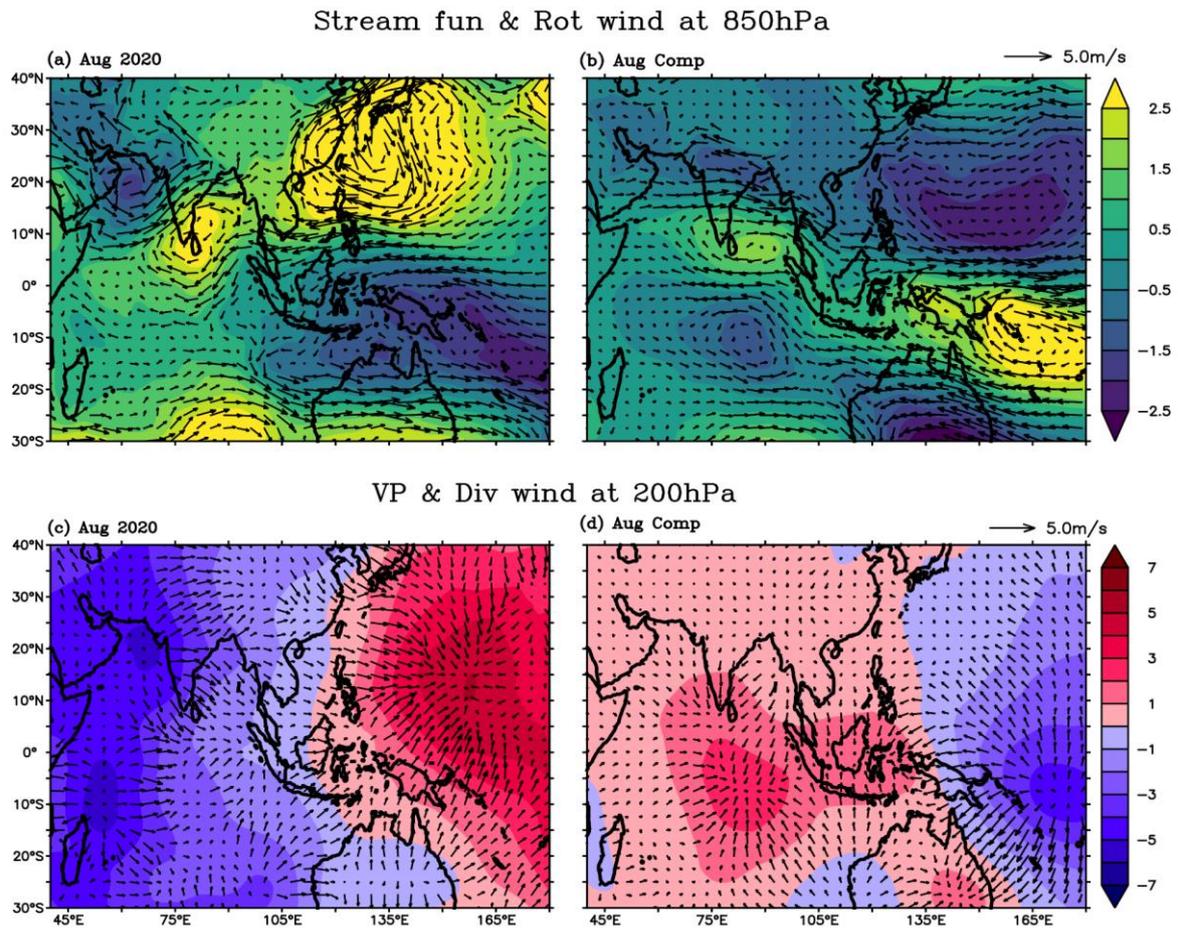

**Figure 9.** (a) 850 hPa Stream function shaded and rotational winds (m/s, vectors)overlayed for August 2020, (b) same as for August Composites. (c) and (d) 200 hPa velocity potential same as (a) and (b).

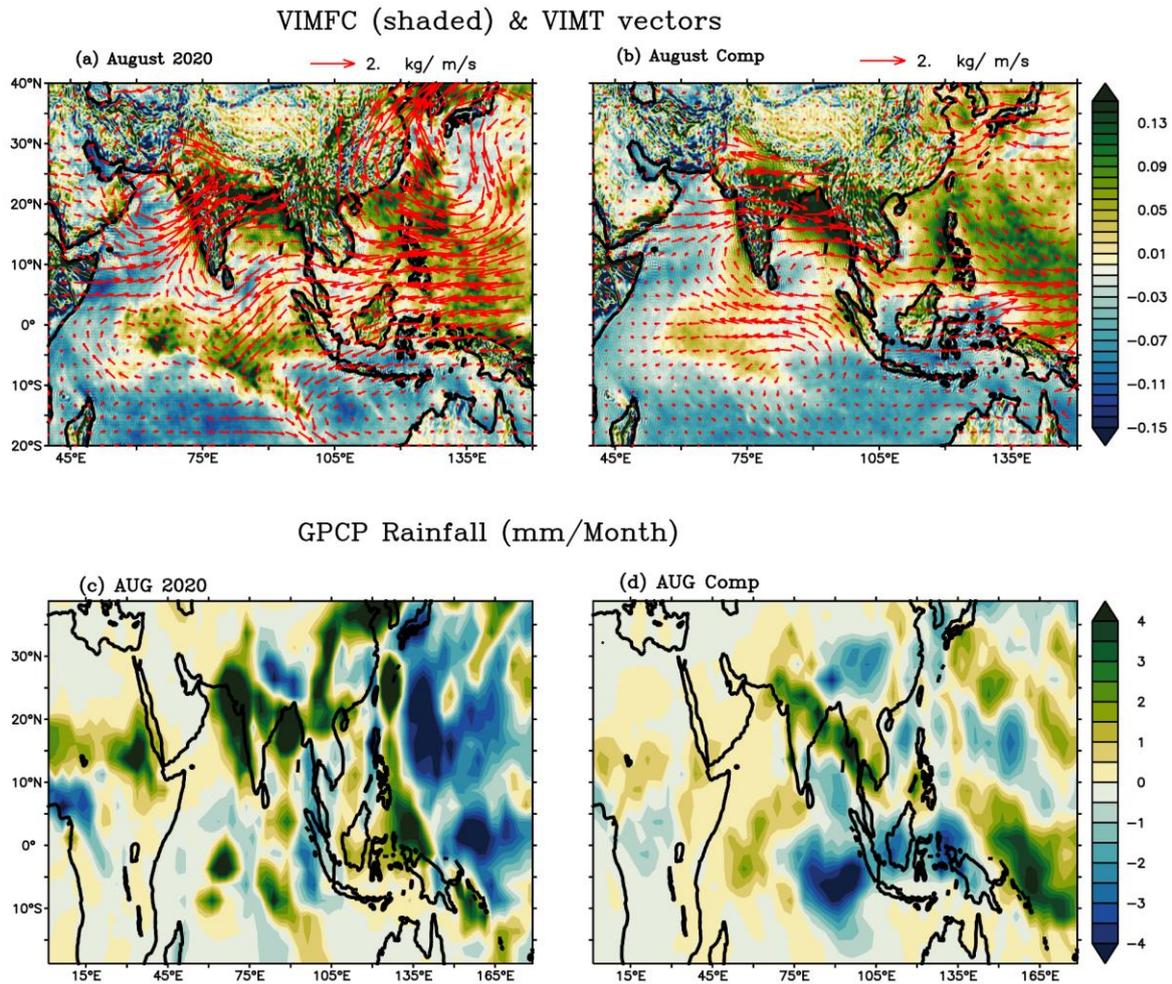

**Figure 10.** (a) Vertically integrated moisture transport for August 2020 (1000–500 hPa levels) and (b) same as (a) but for August composite. The Vectors denote vertically integrated moisture flux transport (kg m$^{-1}$s$^{-1}$) and the moisture flux convergence shaded (kg m$^{-2}$s$^{-1}$). Anomalies of spatial rainfall (GPCP) distribution for August 2020 and August Composite are displayed in (c and d) respectively.

Also, upper-level convergence is evident over the SETIO velocity potential in response to local cooling. This is suggests that enhanced WNP anticyclone reinforces the westward extension of the S-BOB anticyclone which determined the westward shift of convection. A clear westward shift of convection was observed in the vertically integrated moisture convergence as well as transport in August 2020 which is much deep into the head BOB along the monsoon trough in the composites. This is associated with strong cross-equatorial mean flow from SETIO due to prevailing cold SST conditions over that region (e.g fig 10 a and b) and which is consistent with previous studies (e.g., Ashok et al. 2001; Krishnan et al. 2006; Ajayamohan and Rao 2008).The northwesterly vertically integrated moisture transport is visible (e.g fig 10 a and b) during August 2020 from the parts of central Asia, it enhances the rainfall over the NW and central India by enriching the mean monsoon circulation (e.g., Yadav 2017). The anomalous positive (negative) rainfall distribution is evident over the regions with moisture convergence (divergence) over NW & WG (WNP) (fig 10c and d).

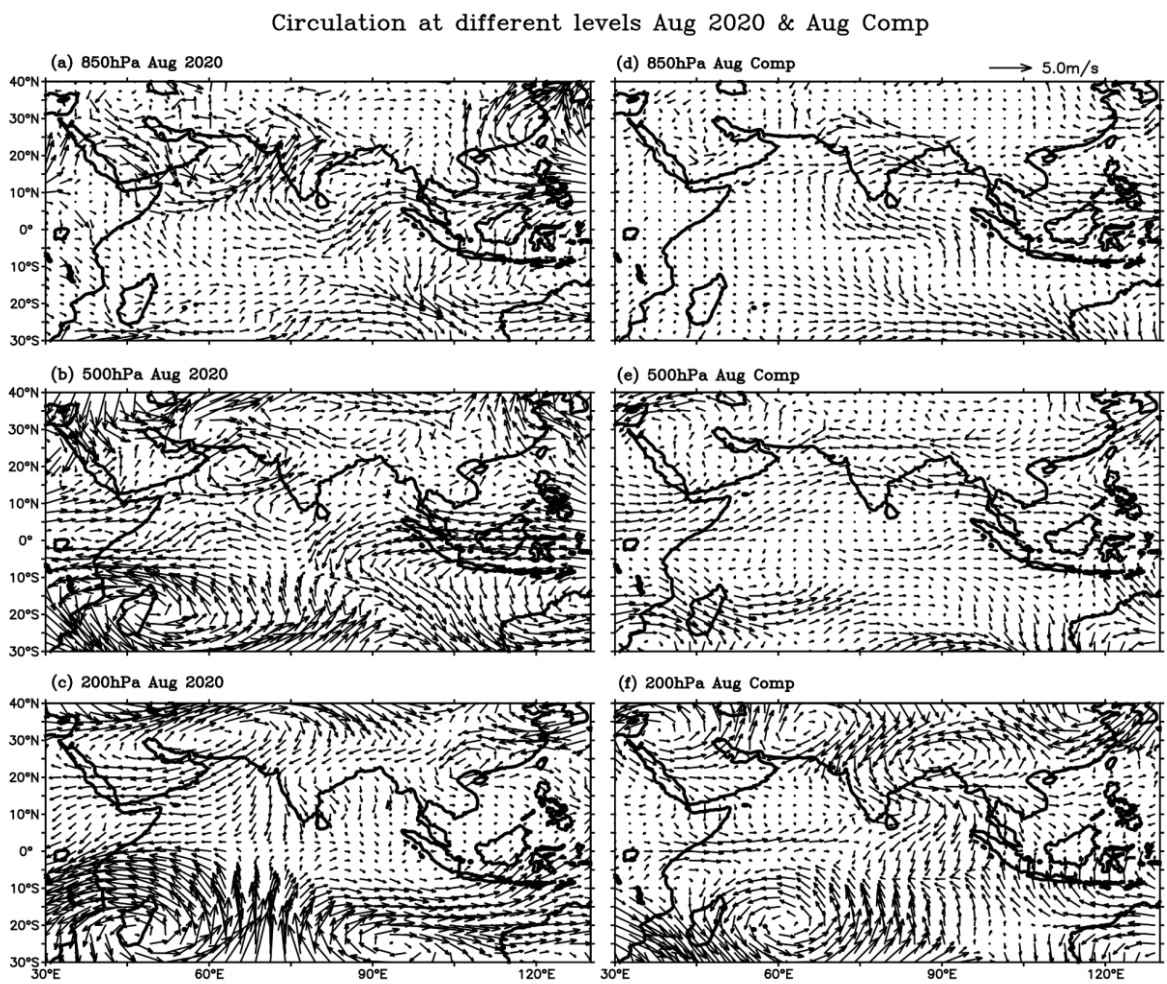

**Figure 11.** Circulation (m/s) at different levels are 850, 500, and 200 hPa, left side panel showing August 2020 (a, b, and c) and the right-side panel for August Composites (d, e, and f).

Composites clearly show strong cross-equatorial flow extending from the SETIO to head BOB. At the same time the cyclonic circulation over the head BOB accompanied by an anticyclone over northwestern parts of Sumatra Islands is evident in observations up to 500 hPa (fig 11 e); Whereas, in August 2020, as mentioned previously strong monsoon flow is associated with cyclonic and anticyclonic circulation on either side of mean monsoon flow over the NPAS and S-BOB. The Intensified westward shift is seen in the Tibetan high during August 2020 (fig 11b), but it is absent in the August composites. Northeasterlies over the Arabian Sea at 200 hPa in August 2020 (fig 11c) are associated with a westward shift anticyclone consistent with lower-level cyclone over NPAS, which is not true for the August composites. However, prevailing strong northerlies over India on either side of cyclone and anticyclone are noted. A strong Tropical Easterly Jet (TEJ) enhances the rainfall over Indian subcontinent (Naidu et al. 2017) by enhancing the lower level westerlies. Apart from strong TEJ, the strong East Asian Westerly Jet (EAWJ) is seen at midlatitudes (~ 40°N). This meandering motion (fig 11c) brings about crests and troughs with pronounced poleward and equatorward displacements of the warm and cold air masses, which are absent in composites. ENSO is known to have a strong influence on the extratropical circulation during the boreal summer by repositioning the monsoonal heat sources. Warming over WNP resulting from fair weather conditions prevailed due to anticyclonic circulation (e.g., Wang et al., 2021a; Qiao et al. 2021; Zhou et al. 2021; Takaya et al. 2020 ) and resulted in shifting of east Asian monsoon convection. Further, it has been reported that the Tropical heating anomalies force a northern hemisphere extratropical response on the interannual time scale (Ding et al. 2011). The studies by Hoerling and Kumar (2003) and Lau et al. (2005) documented that the Indo–western Pacific SST warming could generate a belt of positive, zonally symmetric, upper-level geopotential height that could induce anomalous warmth and dryness. This suggests that the barotropic Rossby wave pattern is closely associated with Asian summer monsoon rainfall (ASM) variability and is also influenced by concurrent ENSO signals. This leads us to further investigate the Rossby waves' role in August 2020 rainfall enhancement and the absence of monsoon depressions.

### 3.3 Barotropic Rossby wave Influence on monsoon synoptic features in August 2020

Anomalous meridional winds pattern defect of the wave pattern which is well known "Silk Road Pattern" during August 2020 displayed the clear northward shift around the 45° - 50°N latitude. Recent studies noted that the in 2020 the large-amplitude stationary Rossby wavetrain originated over the Western North Atlantic (WNA) (Liu et al., 2022). However, in August 2020 EAWJ moves northward, this northward displacement is mainly due to the response to a high-latitude Rossby wavetrain the northerlies (200hPa) incursion was observed up to 5°N along the west coast of India corresponding with southwest to northeast oriented anticyclone over South Central Asia (SCA) (fig 11a). In the composites, a wave pattern is seen along the 35°-40°N latitudes over SCA. This suggests that under the circumstances of August 2020 the wave remains stationary and even retrogress southwest to northeast oriented anticyclone (fig 12). A significant out-phase relation is observed between CI rainfall and 200 hPa meridional wind during August for the period 1979-2020 (fig 13a) which is consistent with an earlier study carried out by Joseph and Srinivasan (1999). The standard deviation of August meridional wind is evident that there is a phase shift of the Rossby waves between weak and strong monsoon years. There are areas of large standard deviation along the Rossby wave path (fig 13b).

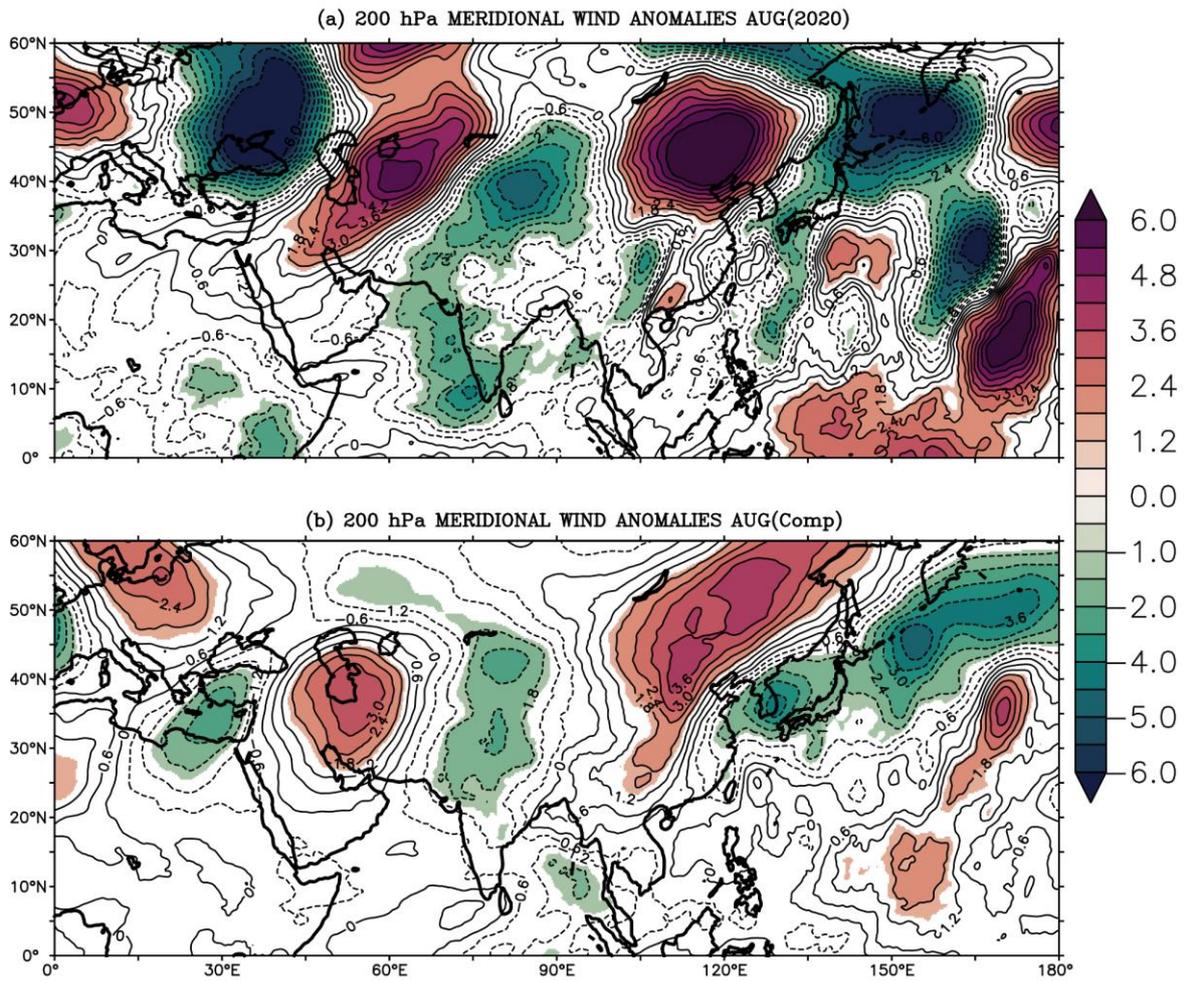

**Figure 12.** (a) 200 hPa meridional wind (m/s) anomalies of August 2020 and (b) August Composites.

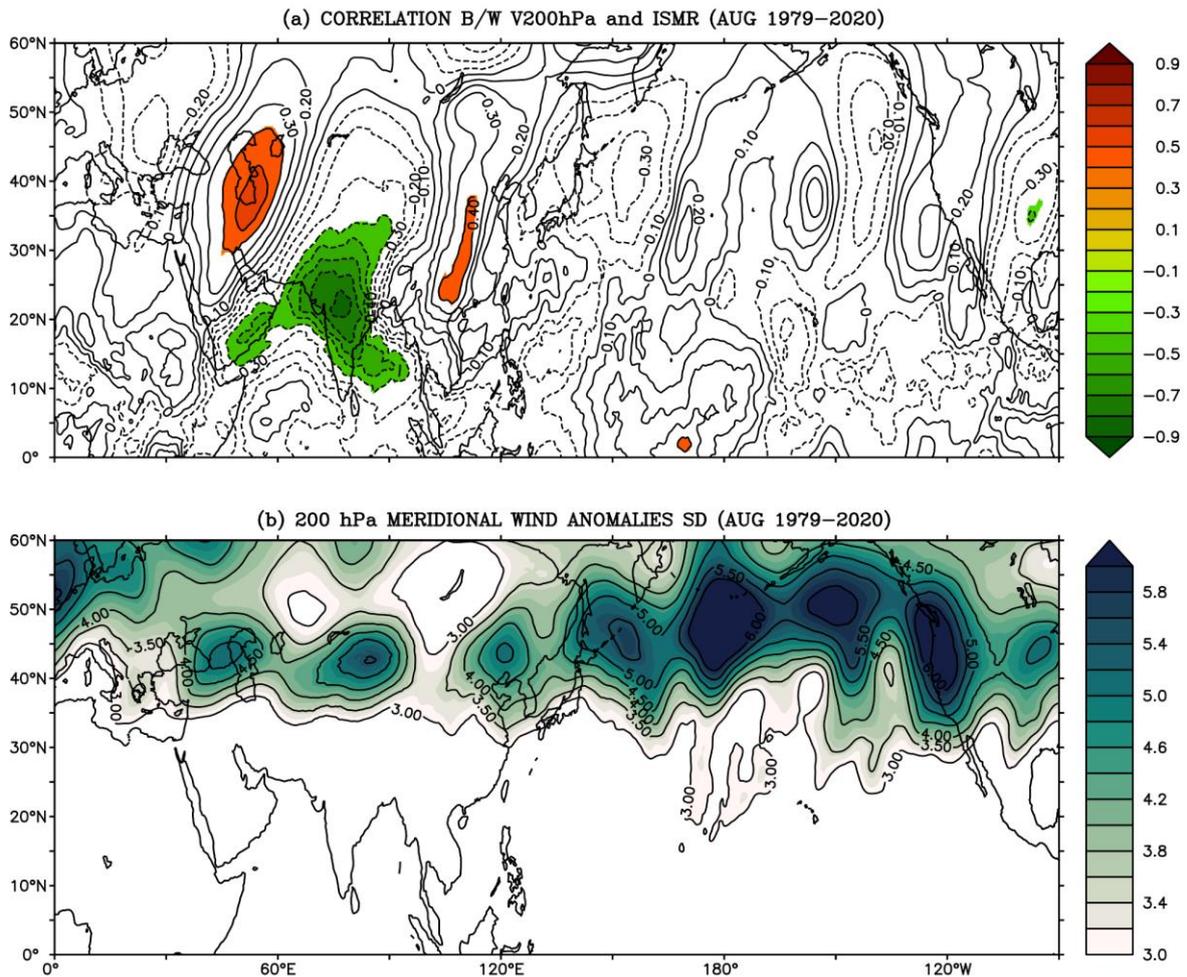

**Figure 13.** (a) Contours showing the Correlation between Indian summer monsoon rainfall and 200 hPa meridional winds for the period of 1979-2020. Shaded significant at 99% confidence level. (b) The standard deviation of August meridional wind for the period of 1979-2020.

This analysis suggests the presence of large-amplitude stationary Rossby waves in the upper atmosphere associated with midlatitude westerlies that trigger the westward intensification of Tibetan anticyclone; fortified monsoon circulation. The Rossby wave activity flux in composites displayed westward (fig 14 b) along mid-latitude around 40°N. In August 2020 a clear southeastward shift in wave activity flux was seen (fig 14 a). It suggested that the upper atmosphere slowly propagated southward and brought cold polar air into the lower latitudes. This accumulates the cold air over SCA due to adiabatic compression, increasing the mid-tropospheric temperature over NW India (fig 7 g) and restricting underlying convection into the lower troposphere.

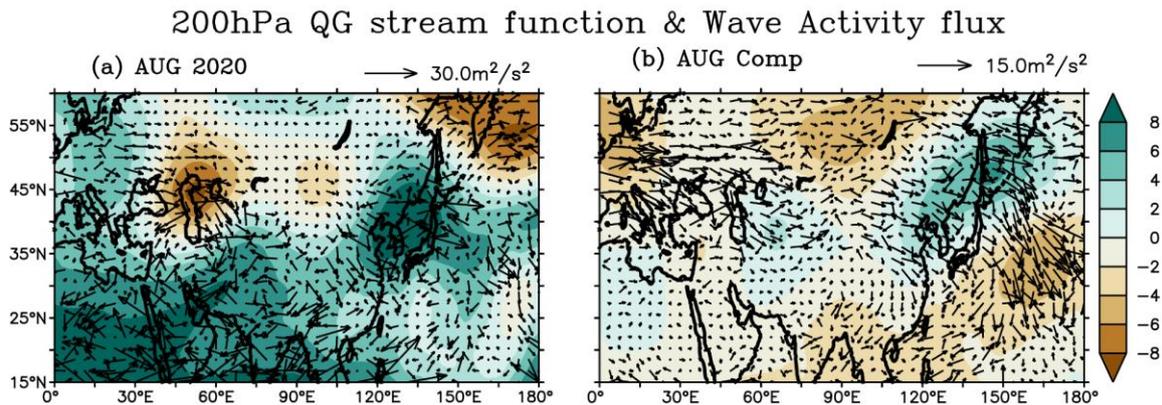

Figure .14 (a) 200 hPa quasi-stationary stream function shaded and Rossby wave activity flux($m^2/s^2$) for August 2020. (b) same as (a) for August composites.

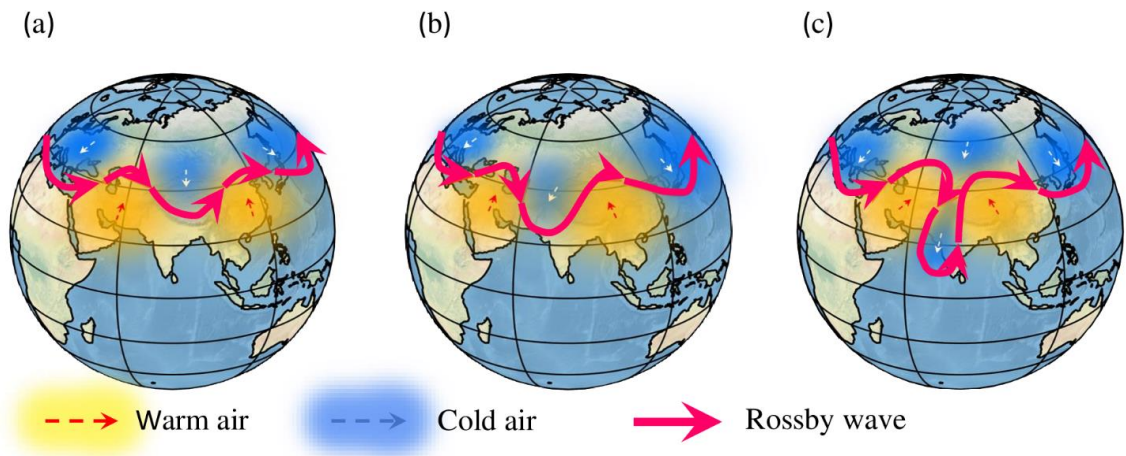

Figure 15. Schematic diagram of Rossby wave during 2020, developing (a, b, and c) and finally detaching a drop of cold air. Orange: warmer air masses; pink: Rossby wave path; blue: colder air masses. This suggests that the Rossby wave under the circumstances of August 2020 remains stationary and retrogresses with NE to SW orientation anticyclone (fig 15 c). It is set on the path of movement of the low-level disturbance beneath and affects rainfall over India, even though lows a not developed into MDs.

We verified the above results in Community Earth System Model Large Ensemble (CESM-LE) historical simulations (Kindly visit the supplementary draft for more detail model analysis). This analysis noted that the CESM-LE show limited skill to simulate the changes in the monsoon rainfall and associated circulation and completely failed to capture the mid-latitude circulation impact on ISM rainfall. It has been reported that many models show poor skills in simulating the effect of mid-latitude circulation influencing ISM rainfall (e.g., Chowdary et al. 2014). CESM-LE reasonably captures the spatial patterns of the SST, with very limited skill in terms of magnitude. Associated circulation consistence with SST patterns results shift in the rainfall over Indian subcontinent this suggest that the CESM-LE underestimates the rainfall patterns it might be due to misrepresentation of large-scale circulation, biases in SST, and associated forcing factors.

## 4. Summary and Conclusions

The interplay of different teleconnections and local conditions along with the role of the barotropic Rossby wave on the Indian Summer monsoon rainfall are analysed in this paper. ISM rainfall is significantly modulated by synoptic-scale systems such as MDs and monsoon lows, which form over the BOB and propagate west/northwestwards into the core monsoon zone of central India. The country received above-normal rainfall during the 2020 summer monsoon season especially in the month of August, supported by six LPAs in the same month. Weak La Niña in the Pacific and TIO warming are accompanied by a strong anticyclonic circulation over the WNP, thereby strengthening the trade winds bringing moisture into the western Pacific as well as the East Asian region. In August 2020 anomalous warming over the equatorial TIO and the NPAS resulted in increased convection over respective regions. In addition, during August 2020 strong northwesterly winds emanating from parts of central Asia merged with enhanced cross-equatorial monsoon flow. However, this flow sheared or dissociated into two branches: one extending up to NW India along the monsoon trough and another one diverging into an anticyclone over the S-BOB, reducing the horizontal shear (Barotropic Instability). The strength of the anticyclone over the S-BOB and its westward shift is determined by the WNP anticyclone. As a consequence of the poor barotropic instability over the SBOB, monsoon lows could not develop into the MDs. On the other hand, a large amplitude stationary Rossby wave in the upper atmosphere deepened into midlatitude, by which those midlatitude westerlies triggered the westward intensification of the Tibetan anticyclone which acts as a conduit for a good monsoon. The upper atmosphere slowly propagated eastward (Rossby wave) with a strong anticyclone over SCA the equatorward branch deepens southward which brought cold polar air into the lower latitudes. It accumulates the cold air over NW India due to adiabatic compression, increasing the mid-tropospheric temperature and restricting underlying convection into the lower troposphere. Barotropic Rossby wave in August 2020 remains stationary and retrogresses with NE to SW orientation anticyclone. It determines the path of movement of the low-level disturbance beneath and affects all India rainfall by elevating the rainfall over NW & WG regions even though lows are not intensified into MDs. Through, CESM-LE model analysis found that the models have limited skill to simulate the changes in the monsoon rainfall and associated circulation and completely failed to capture the mid-latitude circulation impact on ISM rainfall it might be due to misrepresentation of large-scale circulation, biases in SST, and associated forcing factors. Therefore, the relative role of the warm Indian Ocean especially in NPAS and the La Niña development is an open question and we will certainly consider this is for our future work with CESM 2.2 model sensitive experiments.